\documentclass[a4paper,11pt]{article}

\pdfoutput=1
\usepackage{amsfonts}

\usepackage{graphicx}
\usepackage{amsmath}
\usepackage{amssymb}
\usepackage{latexsym}
\usepackage[colorlinks]{hyperref}
\usepackage{color}
\usepackage{float}
\usepackage{cite}
\usepackage{makeidx}
\usepackage{colortbl}
\usepackage{csquotes}
\usepackage[squaren]{SIunits}

\usepackage[textfont={footnotesize,sf},labelfont={color=blue,bf,sf},labelsep=endash]{caption}
\usepackage[position=top,labelfont={color=blue,bf,sf}]{subfig}
\usepackage{bm}

\newcommand{\bra}{\begin{array}}
\newcommand{\era}{\end{array}}
\newcommand{\beq}{\begin{equation}}
\newcommand{\eeq}{\end{equation}}
\newcommand{\bqr}{\begin{eqnarray}}
\newcommand{\eqr}{\end{eqnarray}}

\def\BC{\bb C}
\def\_\BC{\bbi C}

\def\Tr {{\rm Tr}}

\def\( {\left(}
\def\) {\right)}
\def\no2 {{\textstyle{n\over 2}}}

\def\Tr {{\rm Tr}}

\newcommand{\om}{\omega}

\newcommand{\pa}{\partial}

\newcommand{\del}{\delta}
\newcommand{\lga}{\longrightarrow}

\newcommand{\lb}{\label}

\begin{document}

\begin{titlepage}
\setcounter{page}{1}
\renewcommand{\thefootnote}{\fnsymbol{footnote}}

\begin{flushright}
\end{flushright}

\vspace{5mm}
\begin{center}

{\Large \bf {Purity Temperature Dependent for  Coupled  Harmonic Oscillators}}

\vspace{5mm}
{\bf Abdeldjalil Merdaci\footnote{\sf amerdaci@kfu.edu.sa}}$^{a}$,
 {\bf Ahmed Jellal\footnote{\sf 
a.jellal@ucd.ac.ma}}$^{b,c}$, {\bf Ayman Al Sawalha}$^{d}$
and {\bf Abdelhadi Bahaoui}$^{b}$

\vspace{5mm}

{$^a$\em Physics Department, College of Science, King Faisal University,\\
 PO Box
380, Alahsa 31982, Saudi Arabia}


{$^b$\em Saudi Center for Theoretical Physics, Dhahran, Saudi Arabia}

{$^{c}$\em Theoretical Physics Group,  
Faculty of Sciences, Choua\"ib Doukkali University},\\
{\em PO Box 20, 24000 El Jadida, Morocco}

{$^{d}$\em Physics Department,
Faculty of Science,
Jerash University,
Jerash, Jordan}



\vspace{3cm}

\begin{abstract}

We consider the thermal aspect of a system composed of two coupled harmonic oscillators
and study the corresponding purity. 
We initially consider a situation where the system is brought to a
canonical thermal equilibrium with a heat-bath at temperature
$T$.
We adopt the path integral approach
and introduce the evolution operator
to calculate the density matrix and subsequently the  reduced matrix density.
It  is used
to explicitly determine the purity in terms of different
physical quantities and therefore study some limiting cases
related to temperature as well as other parameters.
Different numerical results
are reported and discussed in terms of the involved
parameters of our system.

\vspace{3cm}



\end{abstract}
\end{center}
\end{titlepage}


\section{Introduction}


For a given many body system,
it is not easy
to measure the quantum mechanical
correlations because of their infinitesimal sizes. In same cases for instance
like superconductivity, there exists strong  correlations 
and the pure
states persist even in the large number of the many body state. In such
states 
the classical thermodynamic quantities have been theoretically defined and
identified with measured variables but the quantum mechanical correlations
are lost. One way to recover what were lost is to introduce the so-called
entanglement as a mathematical tool to describe the correlated systems.
Entanglement is one of the most remarkable features of quantum mechanics
that does not have any classical counterpart. It is a notion which has been
initially introduced and coined by Schr\"odinger \cite{1} when quantum
mechanics was still in its early stage of development. Its status has
evolved throughout the decades and has been subjected to significant
changes.

Traditionally, entanglement has been related to the most quantum
mechanical exotic concepts such as Schr\"odinger cat \cite{1},
Einstein-Podolsky-Rosen paradox \cite{2} and violation of Bell's
inequalities \cite{3}. Despite its conventional significance, entanglement
has gained, in the last decades, a renewed interest mainly because  of the
development of the quantum information science \cite{4}. It has been
revealed that it lies at the heart of various  communication and
computational tasks that cannot be implemented classically. It is believed
that entanglement is the main  ingredient of the quantum speed-up in quantum
computation \cite{4}. Moreover, several quantum protocols such as
teleportation,  quantum dense coding, and so on
\cite{Ben1,Ben2,Eckert,Murao,Fuchs,Rausschendorf,Gottesman} are exclusively
realized with the  help of entangled states.

We propose a new approach to explicitly determine the purity function  for a given system
for the whole energy spectrum rather than the ground state as mostly used in the literature. This will
be based on choosing the harmonic oscillators as system and using the path integral technique as tools
to deal with our issues. Our analysis will involve the temperature  as an important parameter that will play a crucial
role in discussing different properties of our system.
To prove the validity of our approach, we will show that our results 
will allow to recover 
the standard case describing the quantum system  in
the ground state at absolute zero temperature, which have been obtained
in our previous work dealing with the 
entanglement in coupled harmonic
oscillators studied using a unitary
transformation
\cite{jellalstat}.

More precisely,
we investigate the entanglement of a system of two coupled harmonic
oscillators~\cite{jellal} by adopting the path integral formalism. Indeed,
we build the propagator corresponding to our system and therefore derive its density 
matrix. This will allow us to determine the wavefunction
describing the state of our system and subsequently end up with the purity
in terms of different physical parameters. We emphasis that
such wavefunction is 
temperature dependent and  associated
to the whole energy spectrum including the ground-state, which offers an exact derivation
of the purity for our system.
We study interesting cases of the
purity by considering different limiting cases related to the temperature, coupling parameter 
and type of particles. At low temperature, we recover our former result obtained in
\cite{jellalstat} and at high temperature the purity shows another behavior.
Jointly to our findings, we present different numerical results and discuss
their basic features.

The present paper is organized as follows. In section 2, we construct the
density matrix of bipartite corresponding to a system of two coupled
harmonic oscillators.
By examining
the two limiting cases of the temperature parameter $\beta$ of the density
matrix and from the well-known solution of the imaginary time Schr\"odinger
equation \cite{Kosztin} we derive the $\beta$-dependent wavefunction, which verifies
the low and high temperature limits as well as
allows to obtain the reduced  density matrix. This will be done by
making use of unitary transformation together with the initial conditions as
well as different changes. In section 3,
from the obtained results we show how to derive the purity in general form
and study interesting cases related to the strengths of different physical
quantities defining our system.
We conclude our results in the final section and give some
perspectives.


\section{Path integral and bipartite}


We consider a system of two coupled harmonic
oscillators of masses $(m_1,m_2)$ parameterized by the planar
coordinates $(x_1,x_2)$. This is governed by a Hamiltonian sum of free and
interacting parts
\begin{equation}  \label{1}
\hat{H}=\frac{\hat{p}_{1}^{2}}{2m_{1}}+\frac{\hat{p}_{2}^{2}}{2m_{2}}+\frac {%
1}{2}C_{1}\hat{x}_{1}^{2}+\frac{1}{2}C_{2}\hat{x}_{2}^{2}+\frac{1}{2}C_{3}%
\hat{x}_{1}\hat{x}_{2}
\end{equation}
where $C_1, C_2$ and $C_3$ are constant parameters. It is clear that the
decoupled harmonic oscillators are recovered by requiring $C_3=0$. In the next, we will
adopt 
the path formalism to explicitly determine the
wavefunction corresponding to the present system and later on study the purity to
characterize the strengths of the entanglement. In doing so, we proceed by
introducing the density matrix and particularly the reduced density matrix.


\subsection{Density matrix}


In the beginning let us establish the mathematical tool that will be used
to attack our concern and deal with different issues. Indeed, in constructing
the the path integral for the propagator corresponding to the  Hamiltonian \eqref{1},
according to \cite{Kosztin, rossi}
we consider the energy shift 
\beq
\hat{H}\longrightarrow \hat{H}-E_{0}\hat{\mathbb I}
\eeq
to ensure that the  wavefunction of the system converges to that
of the ground state at low temperature. Now let us 
  introduce the evolution operator
\begin{equation}
\mathbf{\hat{U}}(\beta)=\mathcal{T}_{D}\exp\left(  -\int_{0}^{\beta}\left(
\hat{H}-E_{0}\hat{\mathbb I}\right)  d\tau\right)  =e^{+\beta E_{0}}\mathcal{T}_{D}%
\exp\left(  -\int_{0}^{\beta}\hat{H}d\tau\right)
\end{equation}
which involves 
the important parameter of our theory that is the temperature $T$ with $\beta=\frac{1}{k_{B}T}$ and the
Boltzmann constant $k_{B}$,  $\mathcal{T}_{D}$ being chronological Dyson
operator. The matrix elements of such propagator take the form
\begin{equation}
\rho^{AB}(x_{1b},x_{2b},x_{1a},x_{2a};\beta)=\langle x_{1b},x_{2b}|\mathbf{%
\hat{U}}(\beta)| x_{1a},x_{2a}\rangle
\end{equation}
where $A$ and $B$ are two subregions forming our system, 
with $\mid x_{1a},x_{2a}\rangle$ and $\mid x_{1b},x_{2b}\rangle$ are the initial
and final states, respectively. 
In the forthcoming analysis, we consider the shorthand notation $%
\rho^{AB}(x_{1b},x_{2b},x_{1a},x_{2a};\beta)=\rho^{AB}\left( b,a;
\beta\right)$. To write this latter in terms of the path integral, one can
divide the temperature parameter $\beta$ into $N+1$ intervals of length $\epsilon =\frac{%
\beta}{N+1}$, use the Trotter formula and insert the completeness relation
\begin{equation}
\int | x_{1},x_{2}\rangle \langle x_{1},x_{2}| dx_{1}dx_{2}=\mathbb{I}
\end{equation}
to get 
the continuous form
\begin{eqnarray}  \label{dmat}
\rho^{AB}\left( b,a; \beta\right) &=& e^{+\beta E_{0}}\\
&& \times \int Dx_{1}Dx_{2}\exp\left\{-
\int_{0}^{\beta}\left( \tfrac{m_{1}}{2}\dot{x}_{1}^{2}+\tfrac{m_{2}}{2}\dot{x%
}_{2}^{2}+\frac{1}{2}C_{1}x_{1}^{2}+\frac{1}{2}C_{2}x_{2}^{2}+\frac {1}{2}%
C_{3}x_{1}x_{2}\right) d\tau\right\} \nonumber
\end{eqnarray}
where the initial conditions 
$x_{1}\left( 0\right) =x_{1a}, x_{2}\left( 0\right) =x_{2a}, x_{1}\left(
\beta\right) =x_{1b}$, $x_{2}\left( \beta\right) =x_{2b}$ 
will be taken into account in order to get the solutions.

To go further in developing the above density matrix, we proceed by
introducing some relevant tools. Indeed, in the beginning we consider the
unitary transformation%
\begin{equation}  \label{tran1}
\left(
\begin{array}{c}
x_{1} \\
x_{2}%
\end{array}
\right) =\left(
\begin{array}{cc}
\frac{1}{\mu}\cos \frac{\theta}{2} & \frac{1}{\mu}\sin \frac{\theta}{2} \\
-\mu\sin\frac{\theta}{2} & \mu\cos \frac{\theta}{2}%
\end{array}
\right) \left(
\begin{array}{c}
X_{1} \\
X_{2}%
\end{array}
\right)
\end{equation}
where we have set the quantities
\begin{equation}\lb{theta}
\tan\theta=\frac{C_{3}}{\mu^{2}C_{2}-\frac{C_{1}}{\mu^{2}}},\qquad
\mu=\left( \frac{m_{1}}{m_{2}}\right) ^{\frac{1}{4}}.
\end{equation}
This transforms the density matrix \eqref{dmat} into the form
\begin{equation}  \label{dens}
\rho^{AB}\left( b,a; \beta\right) =e^{+\beta E_{0}} \int DX_{1}DX_{2}\exp\left\{-
\int_{0}^{\beta}\left( \tfrac{m}{2}\dot{X}_{1}^{2}+\tfrac{m}{2}\dot{X}%
_{2}^{2}+\frac{1}{2}ke^{2\eta}X_{1}^{2}+\frac{1}{2}ke^{-2\eta}X_{2}^{2}%
\right) d\tau\right\}
\end{equation}
and the involved parameters are given by
\begin{equation}\lb{eta}
e^{\pm2\eta}=\frac{\frac{C_{1}}{\mu^{2}}+\mu^{2}C_{2}\mp\sqrt{\left( \frac{%
C_{1}}{\mu^{2}}-\mu^{2}C_{2}\right) ^{2}+C_{3}^{2}}}{2k},\qquad k=\sqrt{%
C_{1}C_{2}-\frac{C_{3}^{2}}{4}}, \qquad m=\sqrt{m_{1}m_{2}}.
\end{equation}
With the above transformation the system is now getting decoupled and
therefore 
\eqref{dens} is separable in terms of the new variables $X_{1}$ and $X_{2}$. Consequently, we have
\begin{equation}
\rho^{AB}(X_{1b},X_{2b},X_{1a},X_{2a};\beta)=e^{+\beta E_{0}} \rho_{1}(X_{1b},X_{1a};\beta
)\rho_{2}(X_{2b},X_{2a};\beta)
\end{equation}
where the two parts read as
\begin{eqnarray}
&&\rho_{1} =\left( \tfrac{m\omega e^\eta}{2\pi\hbar \sinh\left( \hbar\omega\beta
e^\eta\right) }\right) ^{\frac{1}{2}} \exp\left[{\frac{-m\omega e^\eta}{%
2\hbar \sinh\left( \hbar\omega \beta e^\eta\right) }\left( \left(
X_{1b}^{2}+X_{1a}^{2}\right) \cosh\left( \hbar\omega \beta e^\eta\right)
-2X_{1b}X_{1a}\right) }\right] \\
&& \rho_{2} =\left( \tfrac{m\omega e^{-\eta}}{2\pi\hbar \sinh\left( \hbar\omega
\beta e^{-\eta}\right) }\right) ^{\frac{1}{2}} \exp\left[{\frac{-m\omega
e^{-\eta}} {2\hbar \sinh\left( \hbar\omega\beta e^{-\eta}\right) }\left( \left(
X_{2b}^{2}+X_{2a}^{2}\right) \cosh\left( \hbar\omega\beta e^{-\eta}\right)
-2X_{2b}X_{2a}\right) }\right]
\end{eqnarray}
with the  frequency  $\omega=\sqrt{\frac{k}{m}}$. Note that the main
difference between both parts is the sign of the coupling parameter $\eta$, which will play
a crucial role in the forthcoming analysis. Using the transformation \eqref{tran1} to map
the density matrix of the composite system in terms of
the original variables $(x_{1},x_{2})$ as 
\begin{eqnarray}  \label{rhoab1}
\rho^{AB}\left( b,a;\beta\right) & =& \frac{m\omega }{2\pi\hbar } e^{+\beta E_{0}}\left( \tfrac{1} {%
\sinh\left( \hbar\omega\beta e^{\eta}\right) \sinh\left( \hbar\omega\beta
e^{-\eta}\right) }\right) ^{\frac{1}{2}}\exp\left\{
-ax_{1b}^{2}-bx_{2b}^{2}-ax_{1a}^{2}-bx_{2a}^{2}\right\} \\
&& \times\exp\left\{
2cx_{1b}x_{2b}+2cx_{1a}x_{2a}+2dx_{1b}x_{1a}+2fx_{2b}x_{2a}-2gx_{1b}x_{2a}-2gx_{1a}x_{2b}\right\}
\notag
\end{eqnarray}
where different quantities are
\begin{eqnarray}
&&a=\mu^{2}\frac{m\omega }{2\hbar}\left[ {e^{\eta}\coth\left(
\hbar\omega\beta e^{\eta}\right) }\cos^{2} \tfrac{\theta}{2} +{e^{-\eta}\coth\left(
\hbar\omega\beta e^{-\eta}\right) }\sin^{2} \tfrac{\theta}{2} \right] \\
&& b=\frac{m\omega}{\mu^{2}2\hbar}\left[ {e^{\eta}\coth\left(
\hbar\omega\beta e^{\eta}\right) }\sin^{2} \tfrac{\theta}{2} +{e^{-\eta}\coth\left(
\hbar\omega\beta e^{-\eta}\right) }\cos^{2} \tfrac{\theta}{2} \right] \\
&& c=\frac{m\omega}{2\hbar}\left( {e^{\eta}\coth\left(
\hbar\omega\beta e^{\eta}\right) } -{e^{-\eta} \coth\left( \hbar\omega\beta
e^{-\eta}\right) } \right)
\cos \tfrac{\theta}{2} \sin \tfrac{\theta}{2} \\
&& d=\frac{\mu^{2}m\omega}{2\hbar}\left[ \tfrac{e^{\eta}}{\sinh\left(
\hbar\omega\beta e^{\eta}\right) }\cos^{2} \tfrac{\theta}{2} +\tfrac{%
e^{-\eta}}{\sinh\left( \hbar\omega\beta e^{-\eta }\right) }\sin^{2} \tfrac{%
\theta}{2} \right] \\
&& f=\frac{m\omega}{\mu^{2}2\hbar}\left[ \tfrac{e^{\eta}}{\sinh\left(
\hbar\omega\beta e^{\eta}\right) }\sin^{2} \tfrac{\theta}{2} +\tfrac{%
e^{-\eta}}{\sinh\left( \hbar\omega\beta e^{-\eta }\right) }\cos^{2} \tfrac{%
\theta}{2} \right] \\
&& g=\frac{m\omega}{2\hbar}\left( \tfrac{e^{\eta}}{\sinh\left( \hbar
\omega\beta e^{\eta}\right) }-\tfrac{e^{-\eta}}{\sinh\left( \hbar\omega\beta
e^{-\eta}\right) }\right) \cos \tfrac{\theta}{2} \sin \tfrac{\theta}{2}.
\end{eqnarray}
In the next, we will show how \eqref{rhoab1} can be used to deal with different issues related to our system behavior. More precisely,
it will help in getting the corresponding wavefunction describing the whole energy spectrum
of our system 
and later on derive its purity. This  will be used to analyze the degree of
the entanglement and obtain different results. 


\subsection{Reduced density matrix}


To analyze the entanglement of our system, we need first to determine the
reduced density matrix and later evaluate the purity. In doing so, in
early stage we have to derive the wavefunction corresponding to our 
system described by the Hamiltonian \eqref{1}, which has to satisfy
the imaginary time Schr\"odinger equation and gives rise two wavefunctions at 
low and high temperature limits. This will be used to explicitly determine
without approximation the corresponding purity in terms of different physical
parameters. To get such suitable wavefunction, we start 
by making
the variable substitution $\left( x_{1a},x_{2a}\right) =\left(
x_{1b},x_{2b}\right) =\left( x_{1},x_{2}\right)$ into \eqref{rhoab1} and taking only the diagonal elements
of  the density matrix, then we have the probability density
\begin{equation}  \label{eqq20}
P_\beta(x_{1},x_{2})=\mbox{diag} \left(\rho^{AB}(b,a;%
\beta) \right)
\end{equation}
which can be evaluated to get the form
\begin{equation}\lb{gstate}
P_\beta(x_{1},x_{2})= \frac{m\omega e^{+\beta E_{0}}}{2\pi\hbar \sqrt{\sinh\left(
\hbar\omega\beta e^{\eta}\right) \sinh\left( \hbar\omega\beta
e^{-\eta}\right) }} e^{-\tilde{a}(\beta)x_{1}^{2}-\tilde{b}(\beta)x_{2}^{2}+2%
\tilde{c}(\beta)x_{1}x_{2}}
\end{equation}
where we have introduced the shorthand notations 
\begin{align}
\tilde{a}(\beta) & =2(a-d)=\mu^{2} \tfrac{m\omega}{\hbar}\left[
e^{\eta}\tanh\left( \frac{\hbar\omega}{2}\beta e^{\eta}\right) \cos ^{2}
\tfrac{\theta}{2} +e^{-\eta}\tanh\left( \frac{\hbar\omega} {2}\beta
e^{-\eta} \right) \sin^{2} \tfrac{\theta}{2} \right] \\
\tilde{b}(\beta) & =2(b-f)=\tfrac{m\omega}{\mu^{2}\hbar}\left[ e^{\eta}\tanh%
\left( \frac{\hbar\omega}{2}\beta e^{\eta}\right) \sin ^{2} \tfrac{\theta}{2}
+e^{-\eta}\tanh\left( \frac{\hbar\omega}{2}\beta e^{-\eta}\right) \cos^{2}
\tfrac{\theta}{2} \right] \\
\tilde{c} (\beta)& =2(c-g)=\tfrac{m\omega}{\hbar}\left[ e^{\eta}\tanh\left(
\frac{\hbar\omega}{2}\beta e^{\eta}\right) -e^{-\eta}\tanh\left( \frac{%
\hbar\omega}{2}\beta e^{-\eta}\right) \right] \cos \tfrac{\theta}{2} \sin
\tfrac{\theta}{2}.
\end{align}
Note that, \eqref{gstate} looks like the probability density corresponding to the  ground-state of two-dimensional harmonic
oscillator but involving an interesting term that is the last one with $\tilde{c}(\beta)$
parameter. Clearly, if $\tilde{c}(\beta)=0$ we end up with the solution
of decoupled harmonic oscillators.

At this level, let us  check the validity of the probability density \eqref{gstate} by
examining two limiting cases related to the temperature parameter $\beta$. In the
classical limit of  high temperature, the probability distribution takes the form
\begin{eqnarray}\lb{beps}
P_{\beta_0}\left( x_{1},x_{2}\right) &=&
 \underset{\beta\longrightarrow0}{\lim }%
P_{\beta}\left( x_{1},x_{2}\right) \sim\rho_{cl}(x_{1},x_{2};\beta_0)\nonumber\\
&=& \frac{me^{+\beta_0 E_{0}}}{2\pi \hbar^2 \beta_0}
\exp\left[ -\beta_{0} V\left(
x_{1},x_{2}\right) \right] 
\end{eqnarray}
where $V\left( x_{1},x_{2}\right) $ is the potential energy 
\begin{equation}
V\left( x_{1},x_{2}\right) =\frac{1}{2}C_{1}x_{1}^{2}+\frac{1}{2}%
C_{2}x_{2}^{2}+\frac{1}{2}C_{3}x_{1}x_{2}
\end{equation}
and here we have involved the high temperature value $\beta_0$. 
It is clearly seen that $P_{\beta_0}\left( x_{1},x_{2}\right)$
 is nothing but 
the Boltzmann probability distribution 
depending only on the potential energy $V\left( x_{1},x_{2}\right)$. Such result
 is in agreement with the comments of the Kleinert's book \cite{Kleinert}. 
However, in the opposite limit of low temperature
we end up with the the probability distribution associated to the ground-state
\begin{eqnarray}
\underset{\beta\longrightarrow\infty}{\lim}P_{\beta}\left(
x_{1},x_{2}\right)  &\sim&\frac{m\omega}{\hbar\pi}e^{+\beta\left(  E_{0}%
-\hbar\omega\cosh\eta\right)  }e^{-\tfrac{m\omega}{2\hbar}e^{\eta}\left(
\mu\cos\tfrac{\theta}{2}x_{1}-\frac{1}{\mu}\sin\tfrac{\theta}{2}x_{2}\right)
^{2}-\tfrac{m\omega}{2\hbar}e^{-\eta}\left(  \mu\sin\tfrac{\theta}{2}%
x_{1}+\frac{1}{\mu}\cos\tfrac{\theta}{2}x_{2}\right)  ^{2}}\nonumber\\
&=&\left\vert
\psi_{0}(x_{1},x_{2})\right\vert ^{2}
\end{eqnarray}
and it is interesting to stress that such limit of temperature imposes the ground-state energy should be of the form
\beq
E_{0}=\hbar\omega\cosh\eta
\eeq
giving rise the ground-state wavefunction 
\begin{equation}\lb{299}
\psi_{0}(x_{1},x_{2})=\sqrt{\frac{m\omega}{\hbar\pi}} e^{-\tfrac{m\omega}{2\hbar}e^{\eta}\left( \mu \cos
\tfrac{\theta}{2} x_{1}-\frac{1}{\mu}\sin \tfrac{\theta}{2} x_{2}\right)
^{2}-\tfrac{m\omega}{2\hbar}e^{-\eta}\left( \mu\sin \tfrac{\theta}{2} x_{1}+%
\frac{1}{\mu }\cos \tfrac{\theta}{2} x_{2}\right) ^{2}}
\end{equation}
which are in agreement with those obtained in or previous work \cite{jellalstat} just by fixing
the quantum numbers $n_1=n_2=0$.

Generally for any temperature parameter $\beta$,  the wavefunction describing  our system can be determined 
by integrating over the initial variables as has been done in \cite{Kosztin}. Thus, in our case we have to
write the solution of the imaginary time Schr\"odinger equation as
\beq\lb{gfun}
\psi(x_{1},x_{2};\beta) = \int \rho^{AB} \left(b,a;\beta-\frac{\varepsilon}{2}\right) 
\psi\left(x_{1a},x_{2a};\frac{\varepsilon}{2}\right) \ dx_{1a} dx_{2a}
\eeq
where the density matrix of the system verifies the condition
\beq
\underset{\beta\longrightarrow\frac{\varepsilon}{2}}{\lim } \ \rho^{AB}\left(b,a;\beta-\frac{\varepsilon}{2}\right) =
\del(x_1-x_{1a}) \del(x_2-x_{2a}).
\eeq
Note that in order 
to eliminate $\frac{\varepsilon}{2}$ introduced in the initial wavefunction
we have written  the propagator $\rho^{AB}\left(b,a;\beta-\frac{\varepsilon}{2}\right)$
with the prescription $\frac{\varepsilon}{2}$ 
and in this case
\eqref{gfun} is nothing but the exact convolution product.
To go further
it is clear that  one has to fix the form of the initial wavefunction 
$\psi\left(x_{1a},x_{2a};\frac{\varepsilon}{2}\right)$
and then from \eqref{beps} we
choose a non-normalized  initial  wavefunction as
\beq\lb{000}
\psi\left(x_{1a},x_{2a};\frac{\varepsilon}{2}\right)=  
\tfrac{\sqrt
{\tfrac{m\omega}{4\pi\hbar}}}{\sqrt{\cosh\left(  {\hbar\omega e^{\eta}%
}\frac{\varepsilon}{2}\right)  \cosh\left(  {\hbar\omega e^{-\eta}}%
\frac{\varepsilon}{2}\right)  }}
e^{-\frac{1}{2}\tilde{a}(\varepsilon)x_{1}^{2}-\frac{1}{2}\tilde{b}(\varepsilon)x_{2}^{2}
+\tilde{c}(\varepsilon)x_{1}x_{2}}
\eeq
where $\varepsilon$ is a small value of the high temperature, 
which is introduced
to insure 
{the convergence of the probability density  of the initial state.}
Clearly $\psi\left(x_{1a},x_{2a};\frac{\varepsilon}{2}\right)$
does not verify the Schr\"odinger equation but
its square absolute value 
$\left\vert \psi\left(x_{1a},x_{2a};\frac{\varepsilon}{2}\right)\right\vert ^{2}$ is
proportional to the probability density \eqref{beps}.
Now replacing \eqref{000} and integrating \eqref{gfun} to end up with the suitable wavefunction
for the whole energy spectrum
\bqr\lb{333}
\psi(x_{1},x_{2};\beta)  = \tfrac{\sqrt
{\tfrac{m\omega}{4\pi\hbar}}}{\sqrt{\cosh\left(  {\hbar\omega e^{\eta}%
}\beta\right)  \cosh\left(  {\hbar\omega e^{-\eta}}%
\beta\right)  }}
\ e^{+\beta \hbar\omega\cosh\eta}
e^{-\tilde{\alpha} x_{1}^{2}-\tilde{\beta} x_{2}^{2}+2\tilde{\gamma} x_{1}x_{2}}
\eqr
and the quantities take the form
\begin{align}
\tilde{\alpha} & =\mu^{2} \tfrac{m\omega}{2\hbar}\left[
e^{\eta}\tanh\left( {\hbar\omega} e^{\eta} \beta\right) \cos ^{2}
\tfrac{\theta}{2} +e^{-\eta}\tanh\left( {\hbar\omega} 
e^{-\eta} \beta \right) \sin^{2} \tfrac{\theta}{2} \right] \\
\tilde{\beta} & =\tfrac{m\omega}{2\mu^{2}\hbar}\left[ e^{\eta}\tanh%
\left( {\hbar\omega} e^{\eta} \beta \right) \sin ^{2} \tfrac{\theta}{2}
+e^{-\eta}\tanh\left( {\hbar\omega}  e^{-\eta} \beta \right) \cos^{2}
\tfrac{\theta}{2} \right] \\
\tilde{\gamma} & =\tfrac{m\omega}{2\hbar}\left[ e^{\eta}\tanh\left(
{\hbar\omega} e^{\eta} \beta \right) -e^{-\eta}\tanh\left( {%
\hbar\omega}e^{-\eta} \beta \right) \right] \cos \tfrac{\theta}{2} \sin
\tfrac{\theta}{2}.
\end{align}
At this level, we have two interesting comments in order. Indeed, firstly
it is interesting to stress that the obtained wavefunction is temperature dependent and satisfies
the imaginary time Schr\"odinger equation
\beq\lb{schro}
\left(  \hat{H}-E_{0}\hat{\mathbb I}\right) \psi(x_{1},x_{2};\beta) + \frac{\pa}{\pa\beta}\psi(x_{1},x_{2};\beta)=0
\eeq
where 
the substitution $t\lga -i\hbar \beta$ is taken into account. Secondly, in both limiting cases of temperature, $\psi(x_{1},x_{2};\beta)$ \eqref{333}
converges  to the initial wavefunction \eqref{000} at high temperature $\left(\beta\lga \frac{\varepsilon}{2}\right)$ as well as to the 
ground-state wavefunction \eqref{299} at low temperature.

Once the wavefunction corresponding to our system is obtained, we now return
back to explicitly determine the reduced density matrix. 
Then based on the
standard definition
\begin{equation}
\rho_{\mathsf{red}}^{A}(x_{1},x_{1}^{\prime};\beta
)=\frac{\int\psi(x_{1},x_{2};\beta)\psi^{\ast}(x_{1}^{\prime},x_{2}%
;\beta)dx_{2}}{\int\psi(x_{1},x_{2};\beta)\psi^{\ast}(x_{1},x_{2}%
;\beta)dx_{1}dx_{2}}
\label{313}
\end{equation}
we show that the one-particle reduced density matrix takes the form
\begin{equation}  \label{eqred}
\rho_{\mathsf{red}}^{A}(x_{1},x_{1}^{\prime};\beta) =A(\beta) \exp\left( {-\frac{2%
\tilde{\alpha}\tilde{\beta}-\tilde{\gamma}^{2}}
{2\tilde{\beta}}x_{1}^{2}-%
\frac{2\tilde{\alpha}\tilde{\beta}-\tilde{\gamma }^{2}}{2\tilde{\beta}}%
x_{1}^{\prime2}+\frac{\tilde{\gamma}^{2}}{\tilde{\beta}}x_{1}x_{1}^{\prime}}%
\right)
\end{equation}
where the normalization factor $A(\beta)$ verifies the von
Neumann's normalization 
$\Tr_{A}\left(\rho_{\mathsf{red}}^{A}(x_{1},x_{1}^{\prime};\beta)\right)=1$
and is given by
\begin{equation}\label{aeps}
A(\beta)=\sqrt{2\frac{\tilde{\alpha}\tilde{\beta}-\tilde{\gamma}^{2}}
{\pi\tilde {\beta}}}.
\end{equation}
%
We close this part by noting that
in the present analysis
we did not require a normalized wavefunction
to determine the reduced matrix density because such type of function does not satisfy
the imaginary time Schr\"odinger equation \eqref{schro}.


\section{Purity function}


Having obtained the necessary materials, we move to the second step and
derive the general form of the purity function corresponding to our system. 
Later on, we will study interesting limiting cases of such purity in terms of
the strengths of
different physical parameters. 
To underline our system behavior, we will numerically illustrate the obtained
results and give some discussions.


\subsection{General case}


In general, the degree of information about the preparation of a quantum
state 
can be characterized by the associated purity. In other term, to quantify the mixedness of
a state $\rho$ one can use the purity \cite{Adesso, McHugh}
\begin{equation}  \label{defpu}
P= \Tr_{A}\left(\rho\right) ^{2}
\end{equation}
where for $d$-dimensional systems $P$ ranges from $1/d$ for completely mixed
states to 1 for pure states. It is closely related to the linear entropy
measure of mixedness \cite{Wei}
\begin{equation}
S= \frac{d}{d-1} \left(1-P\right)
\end{equation}
which ranges from 0 (for a pure state) to 1 (for a maximally
mixed state). The linear entropy is generally a simpler quantity
to calculate than the von Neumann entropy as there is no
need for diagonalization.

Before determining the purity,
let us first stress that the reduced density matrix  $\rho_{\mathsf{red}}^{B} 
$ corresponding to subregion $B$ has the same form as  in \eqref{eqred},
which can also be obtained by integrating \eqref{eqq20} over the variable $x_1$. Therefore,
for both subregions $A$ and $B$  we have
the same purity, which will be denoted by $P^{A}=P^{B}=P$. 
It is defined as 
trace over square of the reduced density matrix, such as
\begin{equation}
P=\Tr_{A}\left( \rho_{\mathsf{red}}^{A}(x_{1},x_{1}^{\prime};\beta)\right) ^{2}
\end{equation}
or equivalently by using \eqref{eqred} together with \eqref{aeps} 
we obtain
\begin{eqnarray}
P & =&\int\rho_{\mathsf{red}}^{A}(x_{1},x_{1}^{\prime};\beta)\rho_{\mathsf{%
red}}^{A}(x_{1}^{\prime},x_{1};\beta )dx_{1}dx_{1}^{\prime}  \notag \\
& =&A^{2}(\beta)\int e^{-\frac{2\tilde\alpha\tilde\beta-\tilde\gamma^{2}}{\tilde\beta}x_{1}^{2}-2\frac{%
2\tilde\alpha\tilde\beta-\tilde\gamma^{2}}{2\beta}x_{1}^{\prime2}+2\frac{\tilde\gamma^{2}}{\tilde\beta}%
x_{1}x_{1}^{\prime}} dx_1 dx^{\prime}_1.
\end{eqnarray}
After integral performance, we end up with the compact form
\begin{equation}\lb{45}
P=\sqrt{\frac{\tilde\alpha\tilde\beta-\tilde\gamma^{2}}{\tilde\alpha\tilde%
\beta}}
\end{equation}
and also we can replace 
different quantities and make  straightforward calculation to get
the explicit form
%
\begin{eqnarray}  \label{35}
P =\sqrt{\tfrac{\tanh\left( { \hbar\omega
} \beta e^{\eta} \right) 
\tanh\left( { \hbar\omega} \beta e^{-\eta}\right) } 
{ \left(e^{\eta}\tanh\left( { \hbar\omega}
\beta e^{\eta}\right) \sin^{2} \frac{\theta}{2} +e^{-\eta }\tanh\left( {
\hbar\omega} \beta e^{-\eta}\right) \cos^{2} \frac{\theta}{2} \right) \left(
e^{\eta}\tanh\left( { \hbar\omega} \beta e^{\eta}\right) \cos^{2}
\frac{\theta}{2} +e^{-\eta}\tanh\left( { \hbar\omega} \beta
e^{-\eta}\right) \sin^{2} \frac{\theta}{2} \right)}}
\end{eqnarray}
which is showing a strong dependence on the external parameters $\eta$, $%
\theta$ and temperature $\beta$. One more thing, $P$ is actually the product of two quantities and they are
differentiating by the $\eta$ sign of the numerator and the geometric functions in the denominator.
Moreover, we notice that the obtained purity is actually based on exact calculation without use of 
approximation and it is corresponding to the whole energy spectrum.
Consequently to underline our system behavior versus different configurations of the
physical parameters we numerically investigate the basic features of the purity \eqref{35}. 
For the numerical use in the next,  it is convenient to consider the rescaling 
${\hbar \om}\beta \lga \beta$ giving rise  a dimensionless parameter.

\begin{figure}[H]
\centering
  \includegraphics[width=8cm,  height=5.25cm]{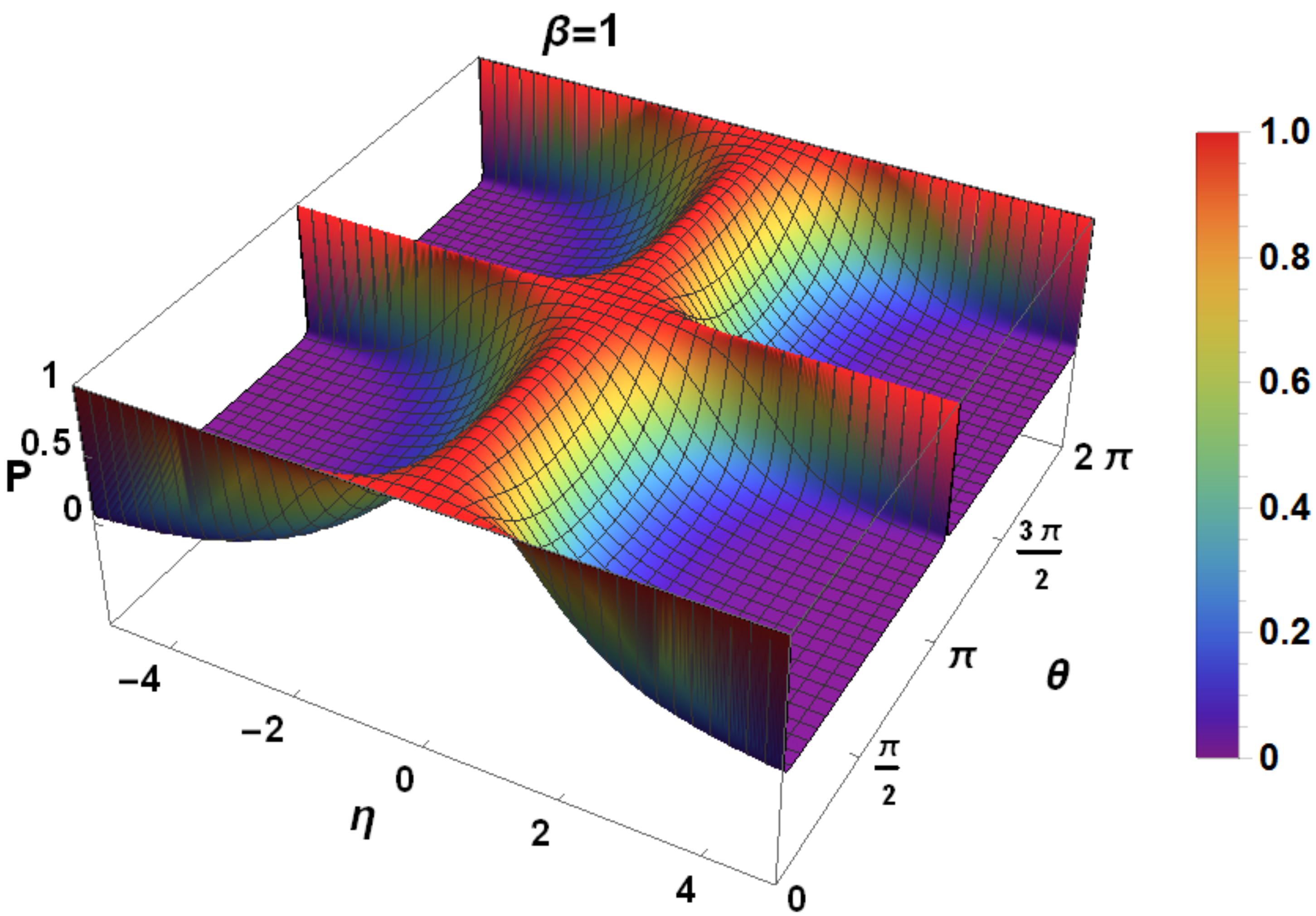}  %
\includegraphics[width=8cm,  height=5.25cm]{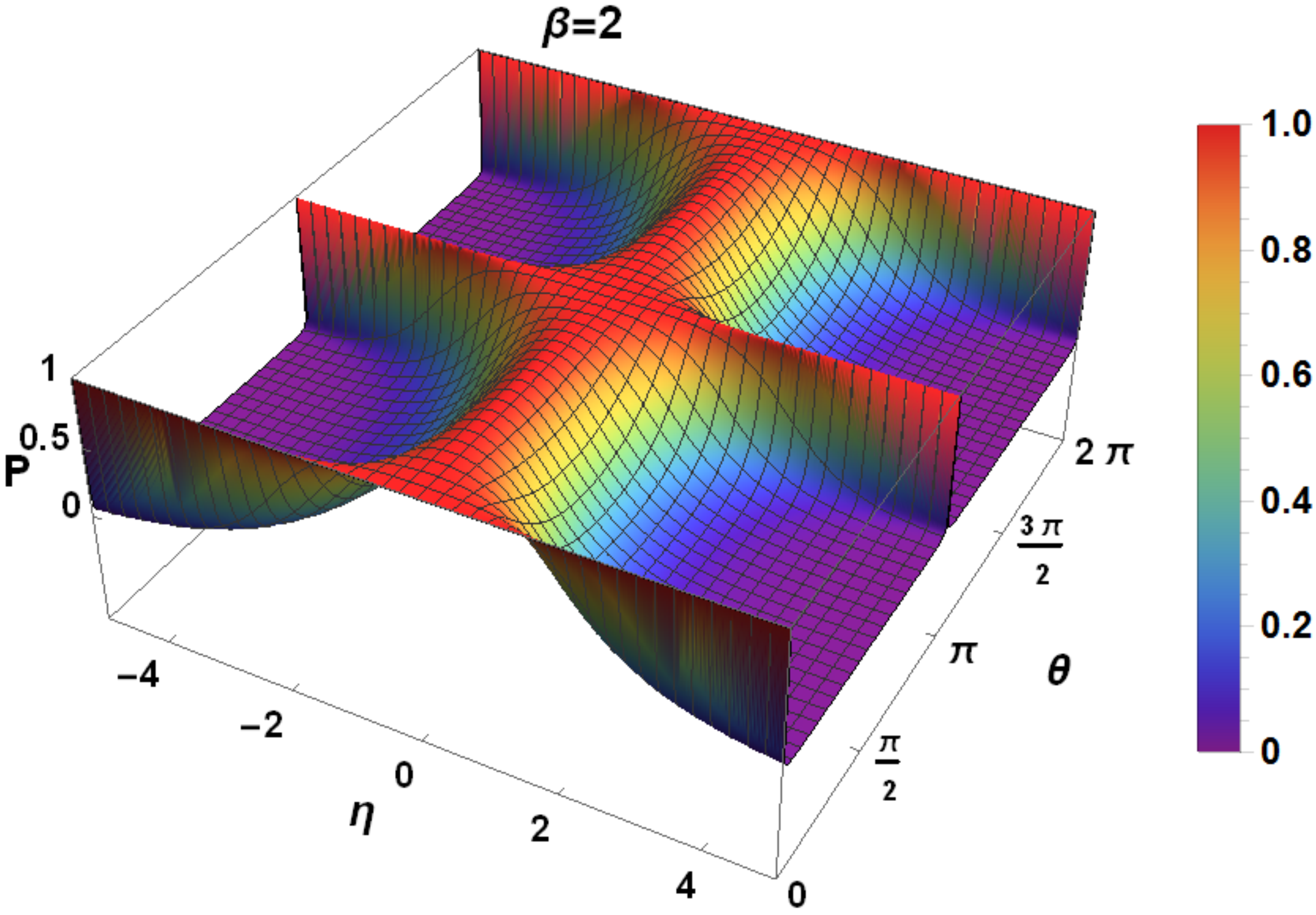}  %
\includegraphics[width=8cm,  height=5.25cm]{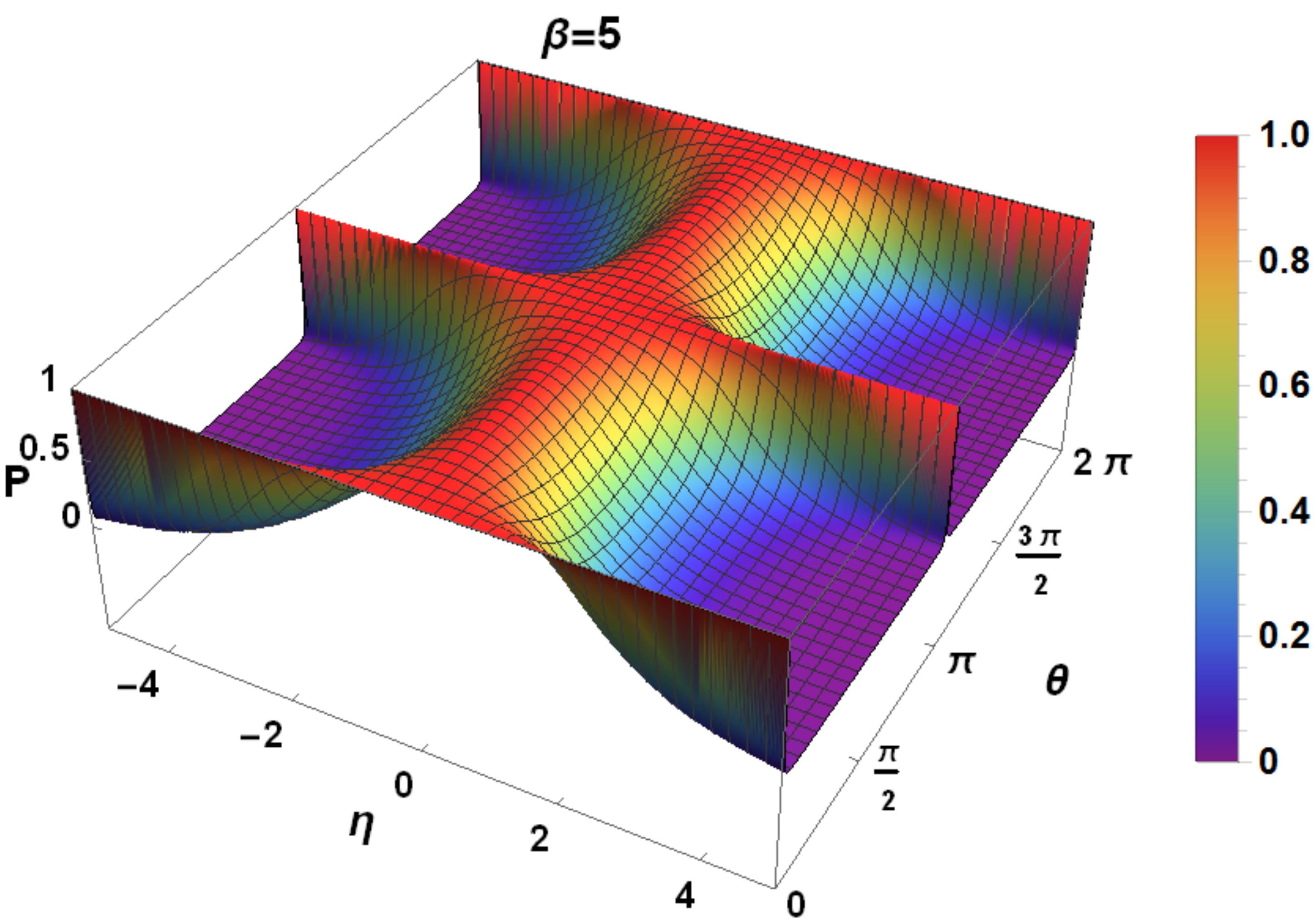}  %
\includegraphics[width=8cm,  height=5.25cm]{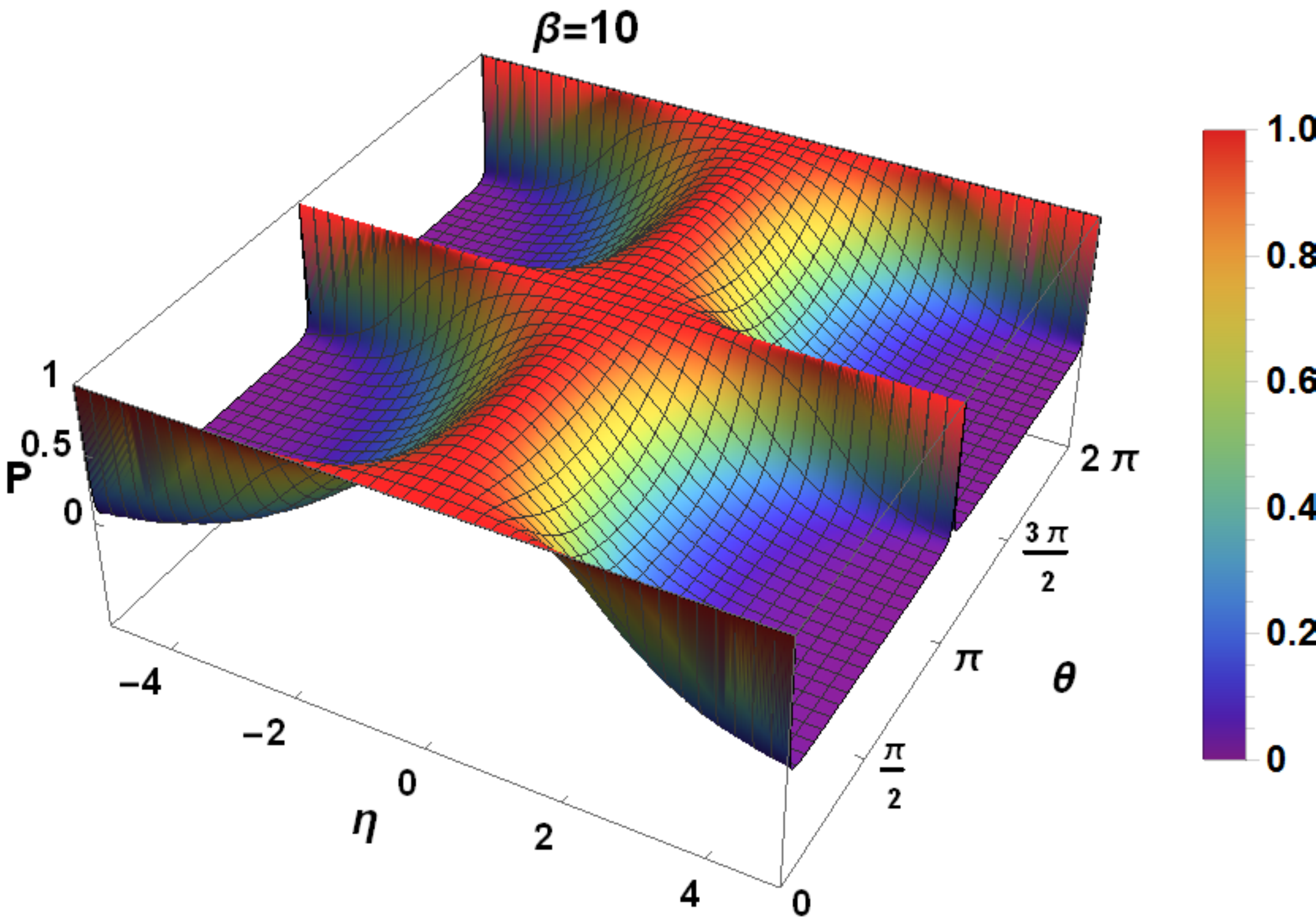}
\caption{\textsf{Purity versus the coupling parameter $\protect\eta$ and the
mixing angle $\protect\theta$ for four fixed value of the temperature $%
\protect\beta=1, 2,5,10$. }}
\label{R35-beta}
\end{figure}

\begin{figure}[H]
\centering  \includegraphics[width=8cm,  height=5.25cm]{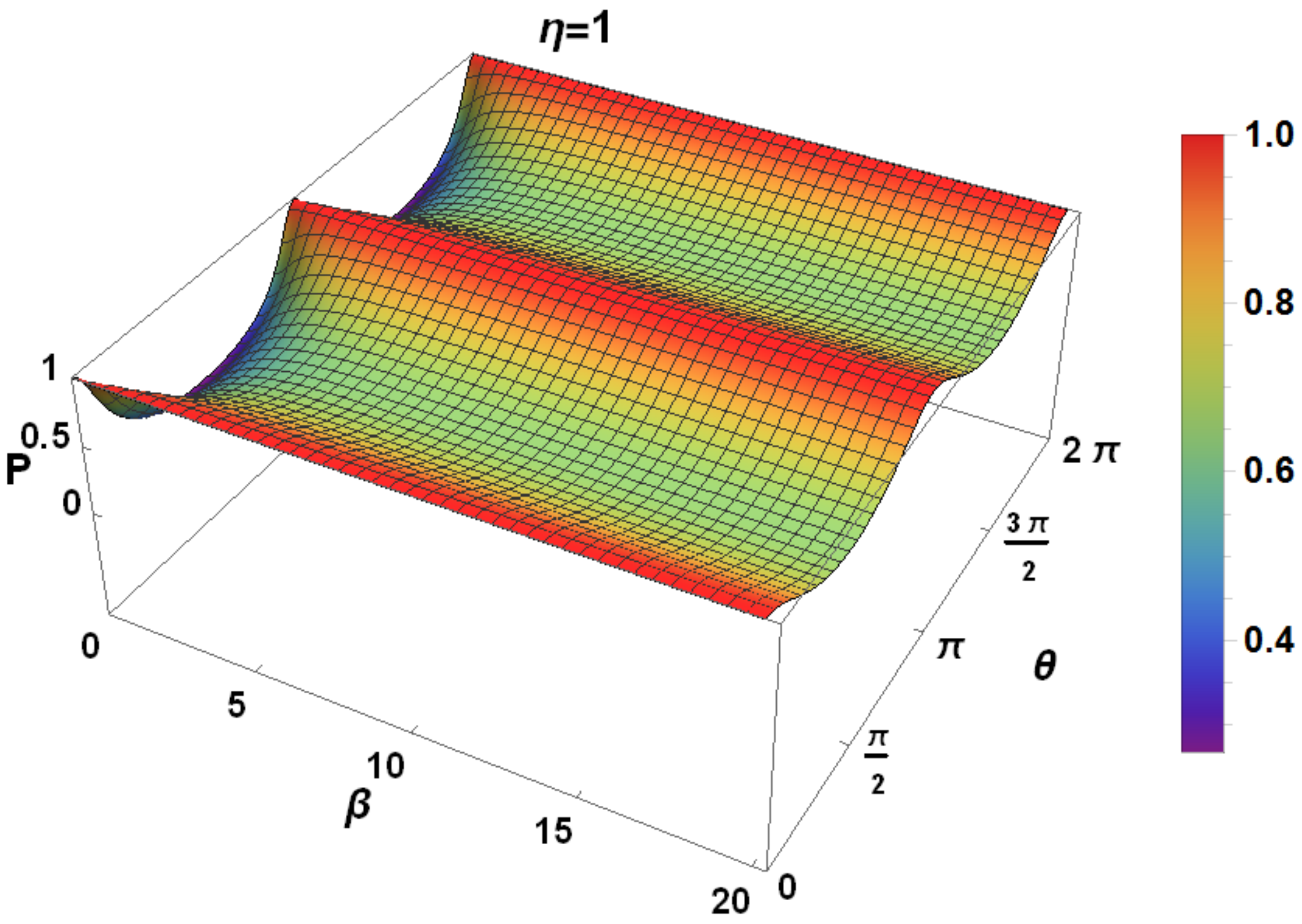}  %
\includegraphics[width=8cm,  height=5.25cm]{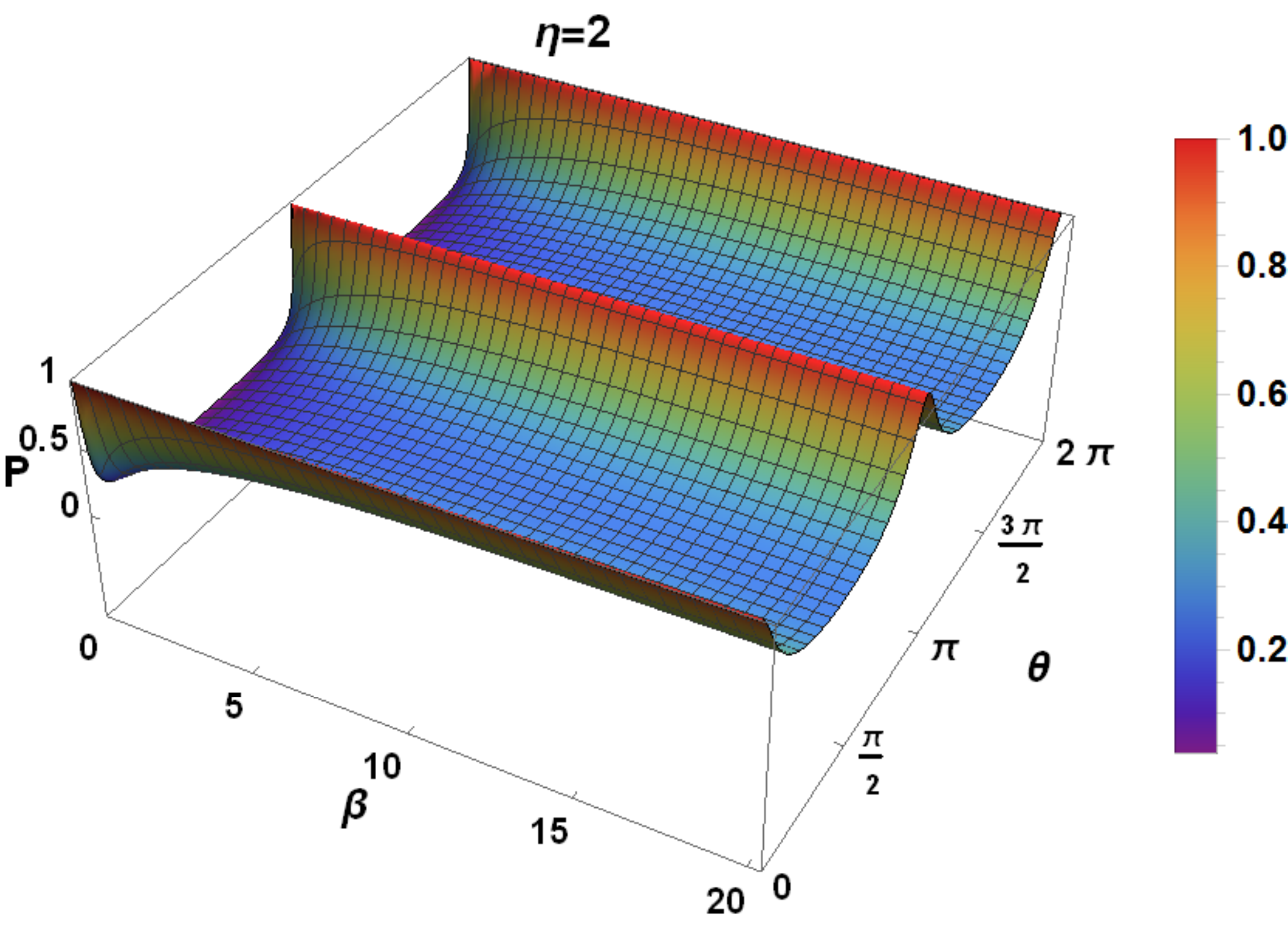} %
\includegraphics[width=8cm,  height=5.25cm]{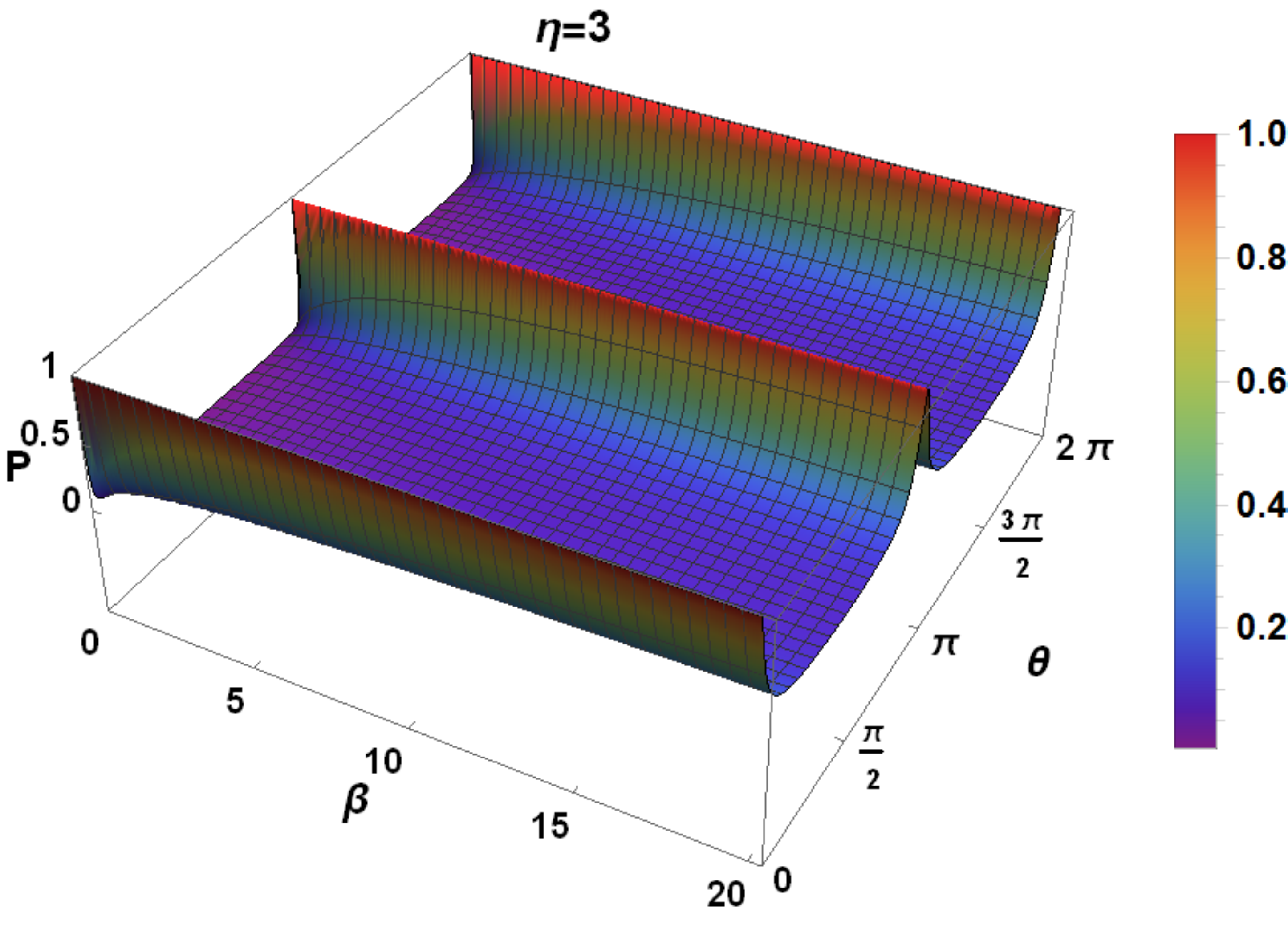}  %
\includegraphics[width=8cm,  height=5.25cm]{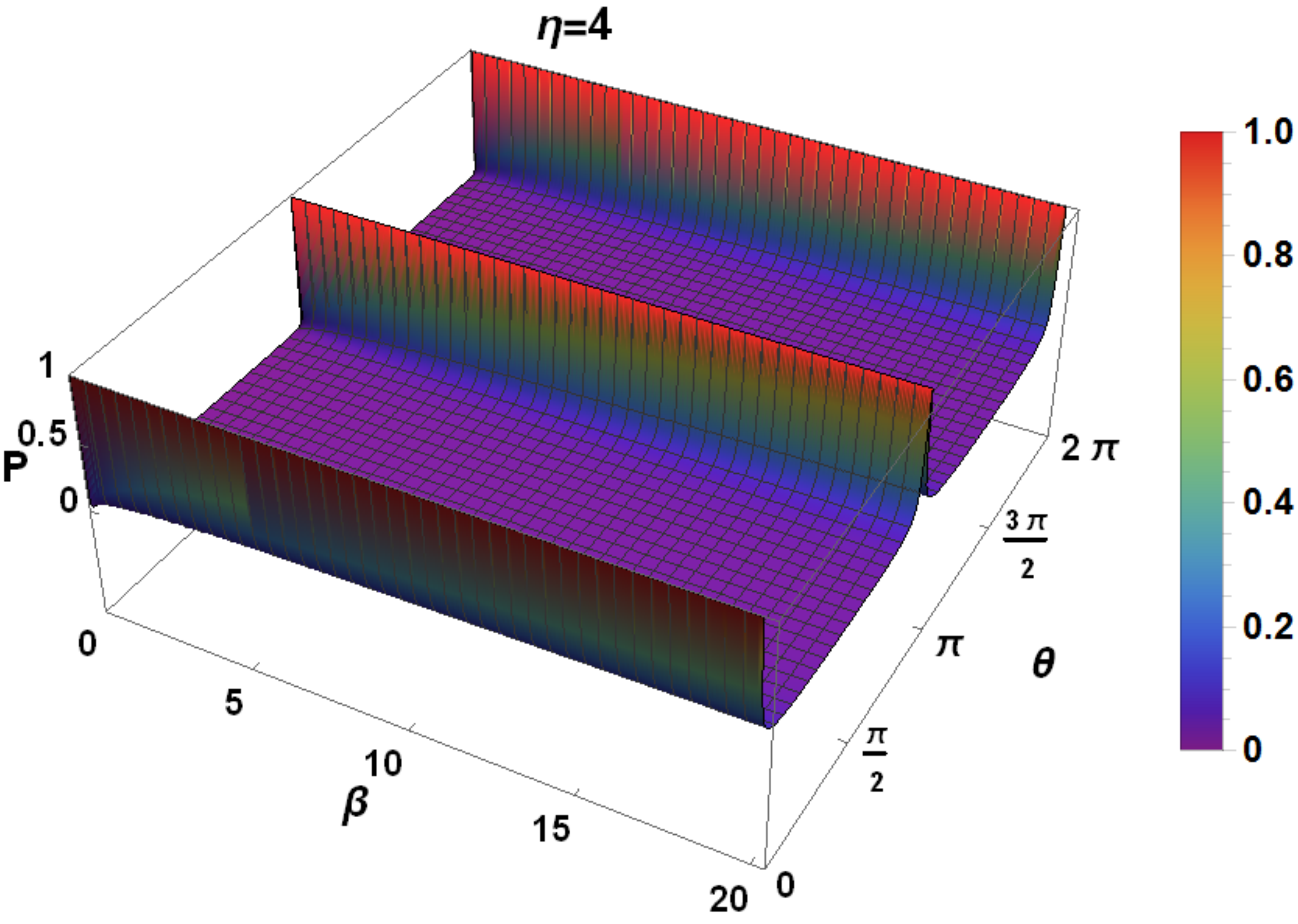}
\caption{\textsf{Purity versus the temperature $\protect\beta$ and the
mixing angle $\protect\theta$ for four fixed value of the coupling
parameter $\protect\eta=1,2,3,4$. }}
\label{R35-eta}
\end{figure}

In Figure \ref{R35-beta}, we plot the purity versus  the coupling parameter $\eta $
and the mixing angle $\theta \in \lbrack 0,2\pi ]$ for four fixed value of
the temperature $\beta =1,2,5,10$. 
{It is clear that the purity, as a function of} $\eta $ and 
$\theta $\, is symmetric with respect to the case $\eta =0$
and the decoupling case $\theta =\pi $ and it lies in the
interval $[0,1]$. It is maximal for $\eta =0$  as well
for $\theta =\pi $, which actually shows that the system is
disentangled. After that it decreases rapidly to reach zero and this
indicates that the entanglement is maximal. More importantly, the purity
becomes constant whenever $\theta $ takes the value zero or $2\pi 
$. This behavior of the purity, 
tells us that one can easily
play with two parameters to control the degree of entanglement in our
 system. As one can see from the four plots that the purity increases as
long as temperature is decreased. 
It means that the temperature plays a great
role in destroying the system entanglement.

Figure \ref{R35-eta} shows the purity versus the temperature $\beta $ and the rotating
angle $\theta $ for four fixed value of the coupling parameter $\eta =1,2,3,4
$. In such situation, we observe that the purity presents another behavior
rather than that observed in Figure \ref{R35-beta}. Indeed,  it  converges rapidly to the minimal and
maximal values as long as the coupling parameter is increased. More
precisely from plot with $\eta =1$ to that with $\eta =4$ there is a radical
changes from highly entangled system to a minimal entanglement.

\begin{figure}[H]
\centering  \includegraphics[width=8cm,  height=5.25cm]{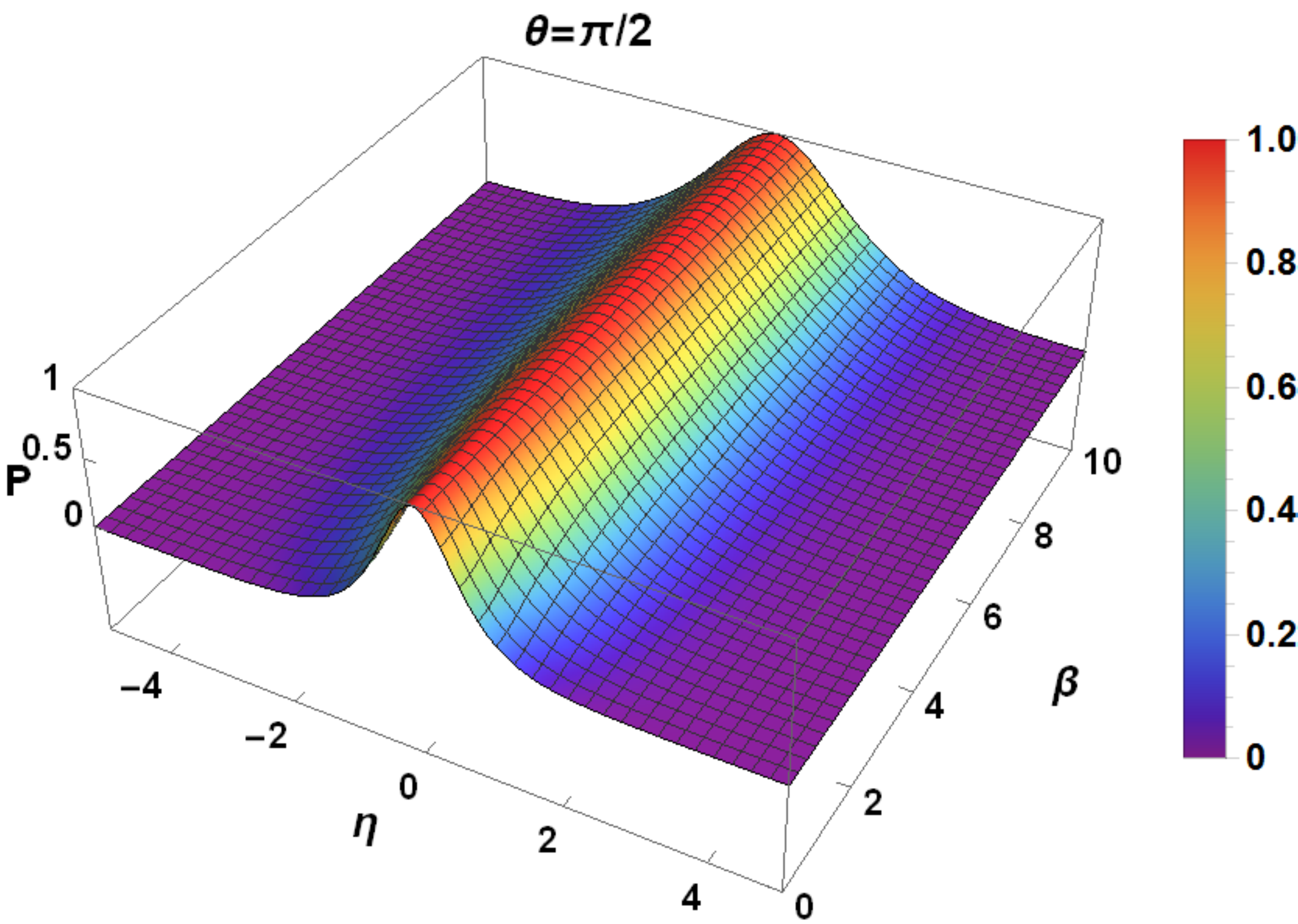}  %
\includegraphics[width=8cm,  height=5.25cm]{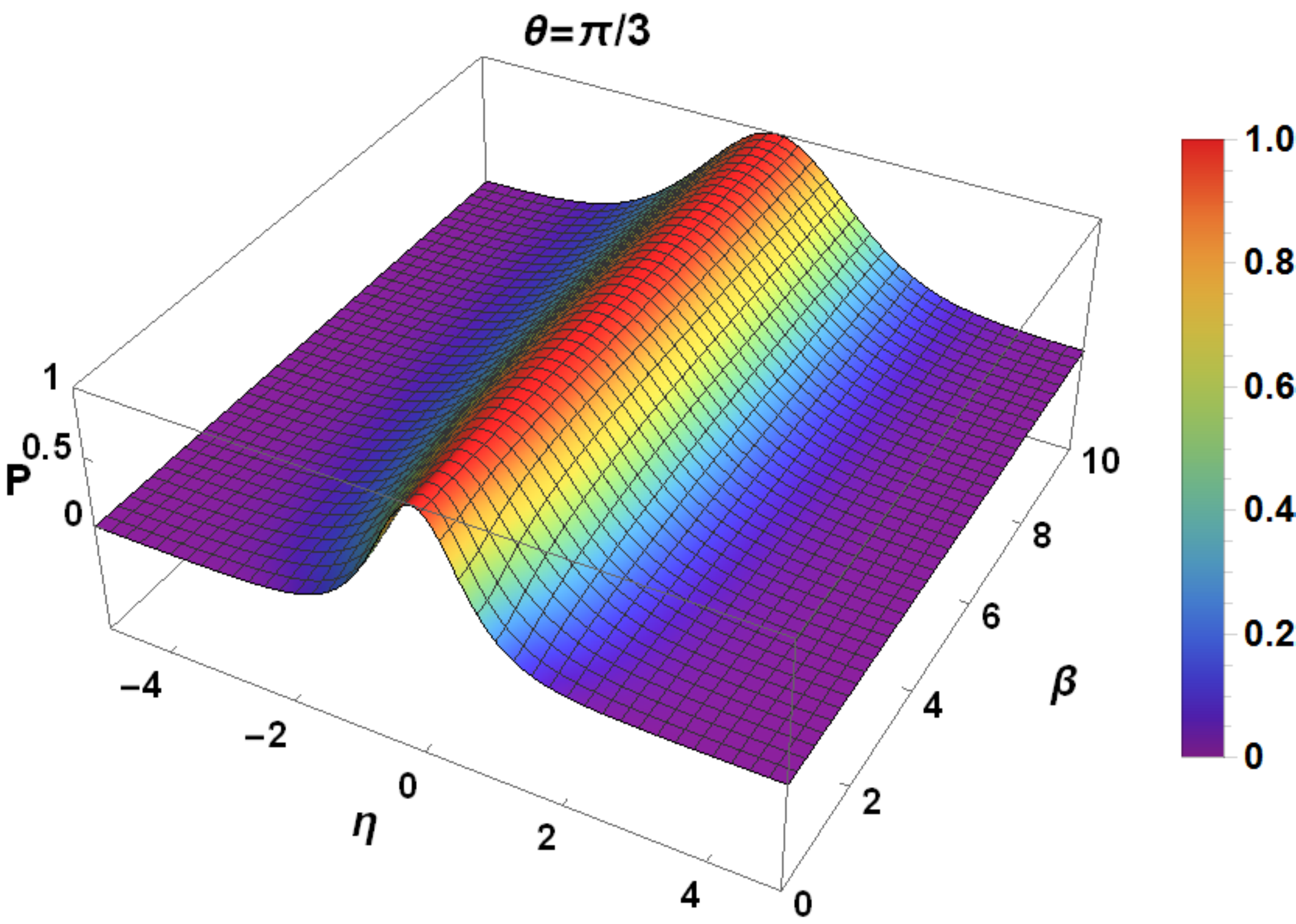}  %
\includegraphics[width=8cm,  height=5.25cm]{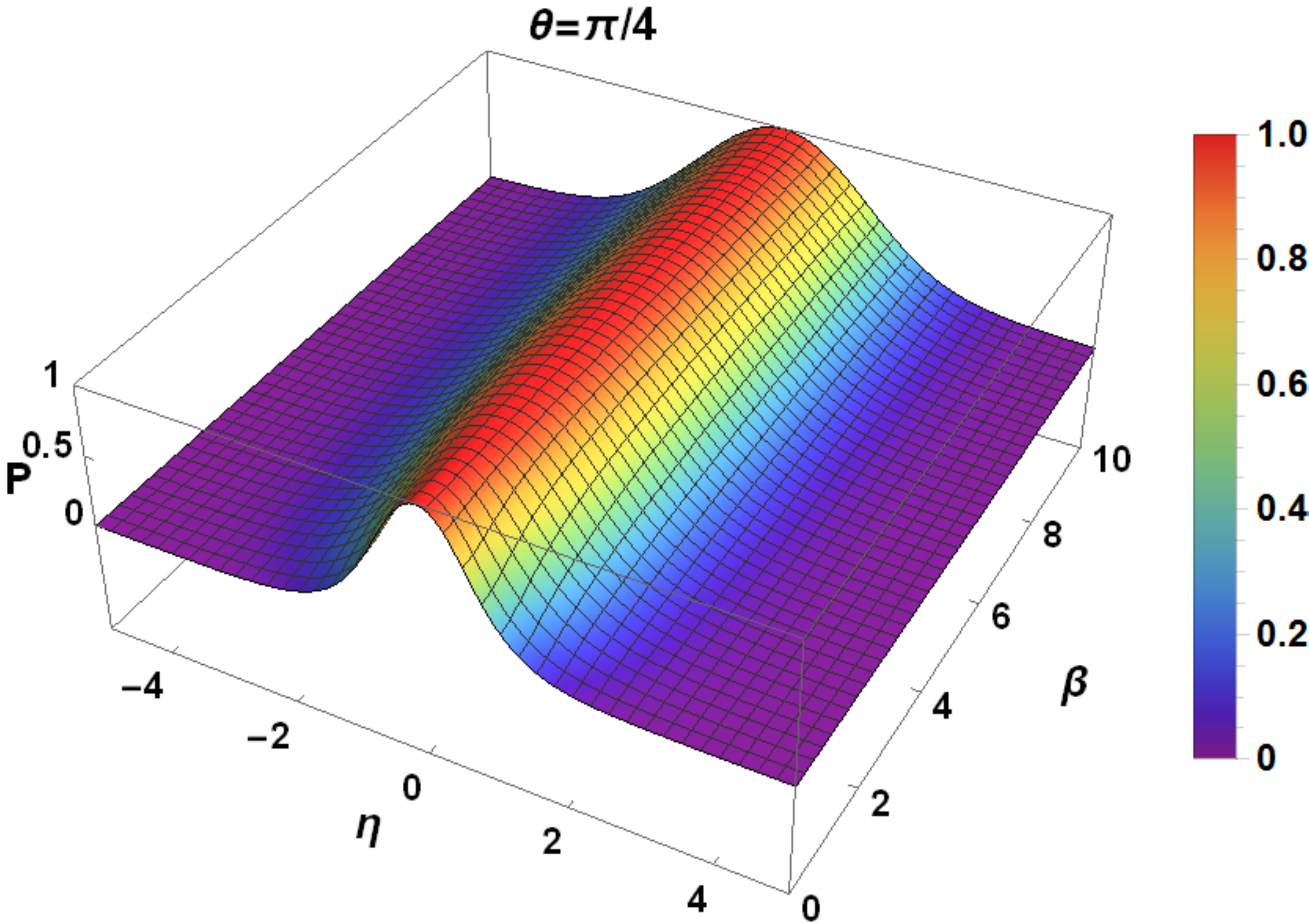}  %
\includegraphics[width=8cm,  height=5.25cm]{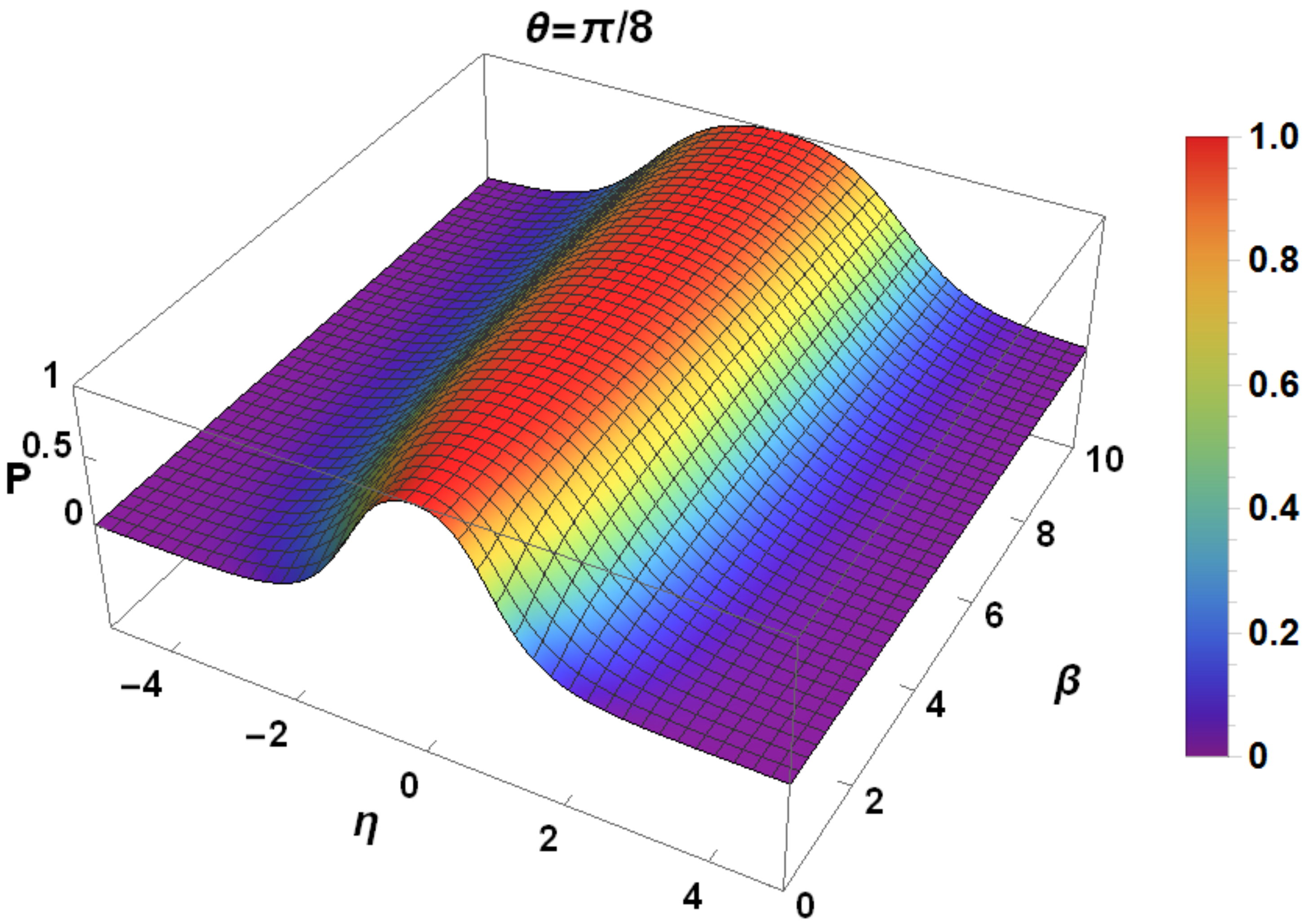}
\caption{\textsf{Purity versus the coupling parameter $\protect\eta$ and the
temperature $\protect\beta$ for four fixed values of the mixing angle $%
\protect\theta=\frac{\protect\pi}{2}, \frac{\protect\pi}{3},\frac{\protect\pi%
}{4},\frac{\protect\pi}{8}$.}}
\label{R35_theta}
\end{figure}

To accomplish such numerical analysis, we inspect the last case by plotting 
the purity versus the coupling parameter $%
\eta $ and the temperature $\beta $ for four fixed values of the rotating
angle $\theta =\frac{\pi }{2},\frac{\pi }{3},\frac{\pi }{4},\frac{\pi }{8}$ in 
Figure \ref{R35_theta}.
It is clearly seen that the purity is increasing slowly from $\theta =\frac{%
\pi }{2}$ to $\theta =\frac{\pi }{8}$, which is showing some minima and maxima.
However the purity takes a maximal value when the coupling parameter $\eta$
is null and also it has a symmetrical behavior.


\subsection{Special cases}


It is interesting to study some liming cases in order to characterize our
system behavior. Since our theory involves three free parameters, one can examine
different situations according to some chosen configurations to simplify the purity function
and therefore extract more information about the entanglement
of our system. These concern the temperature limits, coupling limits and finally
a system composed of identical
particles.


\subsubsection{Temperature limits}


We start our study by looking at the asymptotic behavior
of our system with respect to the temperature parameter $\beta$ limits. Indeed, 
to analyze the low temperature case we take the limit $\beta
\longrightarrow \infty $ in the general form of the purity \eqref{35}. Doing this process 
to get the interesting result
\begin{eqnarray}
P\left( \eta ,\theta ;\beta \longrightarrow \infty \right)  =\tfrac{1}{%
\left|\sin \tfrac{\theta }{2}\cos \tfrac{\theta }{2}\right|\sqrt{\left( 2\cosh \left(
2\eta \right) +\tan ^{2}\tfrac{\theta }{2}+\cot ^{2}\tfrac{\theta }{2}\right)
}} 
\end{eqnarray}%
which ensures
that our system is actually in the ground-state and therefore 
it is exactly the purity $P_{0,0}\left( \eta ,\theta \right) $
corresponding to the ground-state $\left( n_{1}=0,n_{2}=0\right) $ 
obtained in our previous work in dealing with the entanglement
of two coupled harmonic oscillators studied using unitary transformation
\cite{jellalstat}.
This result confirms that
considering the environment temperature as a reservoir acting on our system
is a good candidate to describe its degree of entanglement.

Now let us consider  the high temperature regime and analyze the degree of the entanglement of our  system 
in such situation. This can be worked
out by taking
the limit $\beta\longrightarrow \frac{\varepsilon}{2}$ in \eqref{35} to end up with the purity form
\begin{equation}
P\left( \eta,\theta;\beta\longrightarrow \frac{\varepsilon}{2}\right) =\sqrt{\tfrac{1}{\left(
e^{2\eta}\sin^{2} \frac{\theta}{2} +e^{-2\eta}\cos^{2}
\frac{\theta}{2} \right) \left( e^{2\eta}\cos^{2}\frac{\theta}{%
2} +e^{-2\eta}\sin^{2} \frac{\theta}{2} \right) }}.
\end{equation}
To make comparison with respect to the previous limiting case, we can use  straightforward algebra to rearrange such relation
as
\begin{eqnarray}
P\left( \eta,\theta;\beta\longrightarrow \frac{\varepsilon}{2}\right) &=&
\tfrac{1}{\left|\sin\frac{\theta}{2}%
 \cos\frac{\theta}{2}\right| \sqrt{ 2\cosh\left( 4\eta\right)
+\tan^{2} \frac{\theta}{2} +\cot^{2} \frac{\theta}{2}%
 }}\nonumber \\
 &=&\sqrt{\tfrac{C_{1}C_{2}-\frac{C_{3}^{2}%
}{4}}{C_{1}C_{2}}}.
\end{eqnarray}
It is clear that now we
can establish a link between both of temperature regimes and therefore 
write the following relation
\beq
P\left( \eta,\theta;\beta\longrightarrow \frac{\varepsilon}{2}\right) =P\left( 2\eta,\theta;\beta\longrightarrow\infty\right)
\eeq
which tells us that it is matter of controlling the values taken by the coupling parameter $\eta$ to
go from a regime to another.

\begin{figure}[H]
\centering  \includegraphics[width=12cm,  height=8cm]{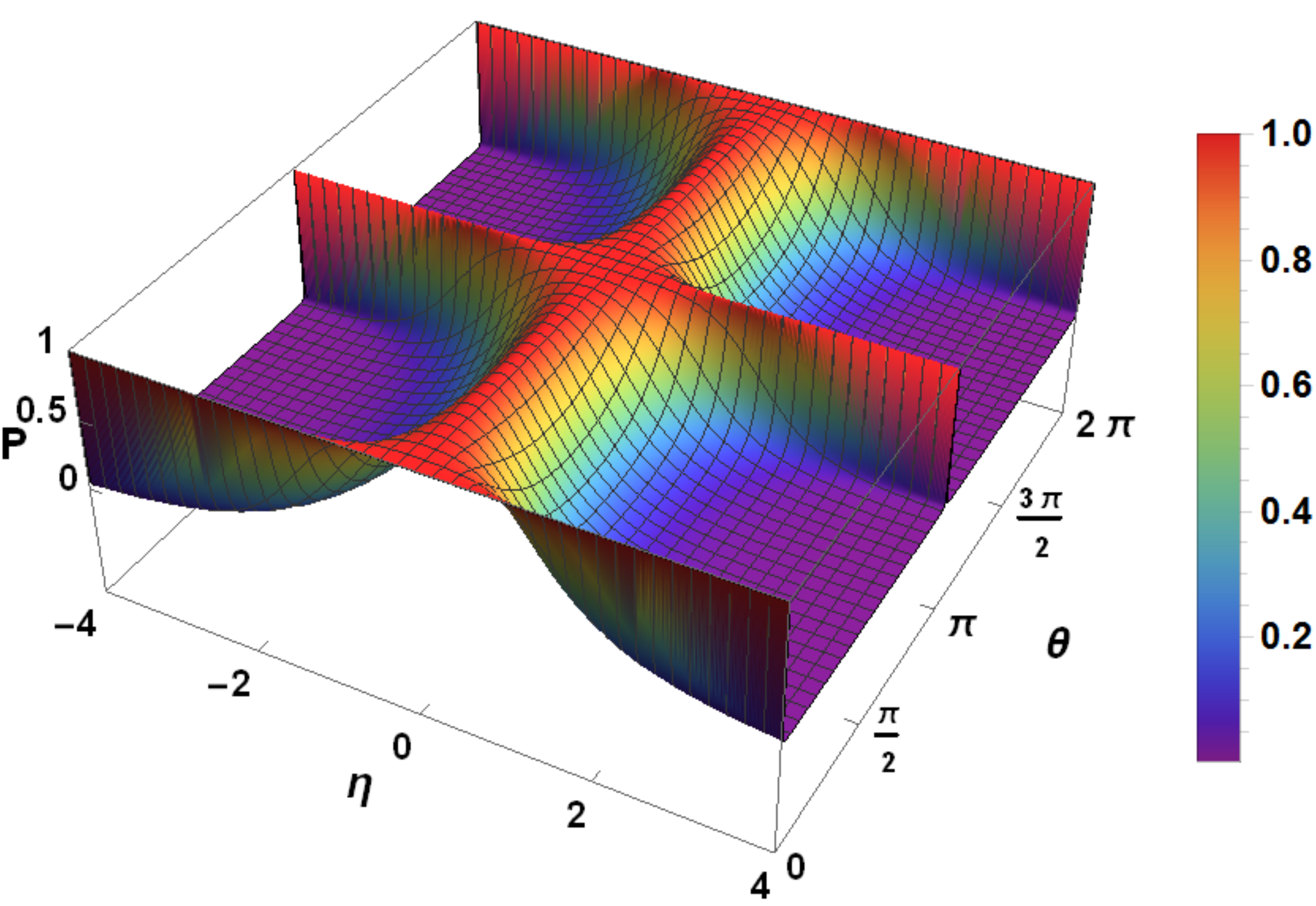}
\caption{\textsf{Purity versus the coupling parameter $\protect\eta$ and the
mixing angle $\protect\theta$ for high temperature limit. }}
\label{R38-2}
\end{figure}

Figure \ref{R38-2}  shows the high temperature regime for the purity as function
of the coupling parameter $\protect\eta$ and the
mixing angle $\protect\theta$. We observe that $P$
increases slowly and at some interval of $\eta $ with $\theta \in \lbrack
0,\pi ]$, it  reaches the maximum where the system becomes highly disentangled
and the same behavior happened for 
$\theta \in \lbrack \pi ,2\pi ]$. Also $P$ shows a symmetrical behavior for
two values $\eta=0$ and $\theta=\pi$, which are corresponding to the maximal value
of $P$.

\subsubsection{Coupling limits}


Two situations will be analyzed with respect to the strength of the
coupling parameter $\eta$, which will allow us to see how much our system is
entangled. 
We start with the weak coupling that is characterized by taking the limit  $%
C_{3}\longrightarrow0$ where the angle $\theta\longrightarrow\theta_{w}$ and
the coupling $\eta\longrightarrow\eta_{w}$. In this case, 
\eqref{theta} and \eqref{eta} reduce to the following
\begin{equation}
e^{2\eta_{w}}=\frac{1}{\mu^{2}}\sqrt{\frac{C_{1}}{C_{2}}},\qquad
\theta_{w}=0
\end{equation}%
which can be implemented into
\eqref{35} to get the maximal value of the purity
\begin{equation}
P\left( \eta_{w},\theta_{w};\beta\right) =1
\end{equation}
and therefore it is showing that our system is completely separable and consequently there is no
entangled states.

Now we consider the strong coupling limit and derive the corresponding
purity. In doing so, we notice that if the limit  $C_{3}\longrightarrow2%
\sqrt{C_{1}C_{2}}$ is required  then one could obtain
\begin{eqnarray}
&& \tan\theta_{s}\longrightarrow \frac{2\sqrt{C_{1}C_{2}}}{\mu^{2}C_{2}-%
\frac{C_{1}}{\mu^{2}}}\longrightarrow0 \\
&& \eta\longrightarrow\eta_{s}=+\infty, \qquad k\longrightarrow0^{+}.
\end{eqnarray}
Combining all to end up with the results
\begin{eqnarray}
ke^{2\eta_{s}} \longrightarrow\frac{C_{1}}{\mu^{2}}+\mu^{2}C_{2}, \qquad
\theta_{s} =\tan^{-1}\left( \frac{2\sqrt{C_{1}C_{2}}}{\mu^{2}C_{2}-\frac{%
C_{1}}{\mu^{2}}}\right)
\end{eqnarray}
and thus the purity \eqref{35} reduces to the following quantity
\begin{equation}
P^{A}\left( \eta_{s},\theta_{s};\beta\right) \longrightarrow0
\end{equation}
which is 
telling us 
that our system is maximally entangled. This summarize that there are two extremely  values of the
purity those could be reached as long as the coupling parameter takes small
or large values.


\subsubsection{Identical particles}


The last situation is related to the nature of our system, 
which is equivalent to require that both of  harmonic oscillators have the same
mass $m_{1}=m_{2}$ and
frequency $C_{1}=C_{2}$. Thus from \eqref{theta} and \eqref{eta}, we end up with the constraint
$\theta\longrightarrow\frac{\pi}{2}$ and $\eta\longrightarrow \eta_{id}$
with
\begin{equation}
e^{2\eta_{id}}=\sqrt{\frac{C_{1}+\frac{C_{3}}{2}}{C_{1}-\frac{C_{3}}{2}}}
\end{equation}
and replacing in the purity \eqref{35} to obtain
\begin{equation}
P\left( \eta_{id},\theta=\frac{\pi}{2};\beta\right) =\tfrac{2\sqrt {%
\tanh\left( \hbar\sqrt{\frac{C_{1}+\frac{C_{3}}{2}}{m_{1}}} {\beta}%
\right) \tanh\left( \hbar\sqrt{\frac{C_{1}-\frac{C_{3}}{2}}{m_{1}}} {%
\beta} \right) }}{\left( \frac{C_{1}+\frac{C_{3}}{2}}{C_{1}-\frac{C_{3}}{2}%
}\right) ^{\frac{1}{4}}\tanh\left( \hbar\sqrt{\frac {C_{1}+\frac{C_{3}}{2}}{%
m_{1}}} {\beta} \right) +\left( \frac {C_{1}-\frac{C_{3}}{2}}{C_{1}+%
\frac{C_{3}}{2}}\right) ^{\frac{1}{4}}\tanh\left( \hbar\sqrt{\frac{C_{1}-%
\frac{C_{3}}{2}}{m_{1}}} {\beta} \right) }
\end{equation}
which is depending on the temperature. 
We notice that there is a special value of the coupling parameter  reducing the above expression to a 
simplified form. Indeed, by requiring
$C_{3}=\frac{30}{17}C_{1}$, we obtain  $ \eta_{id}=\ln2$ and
and therefore the purity becomes
\begin{equation}
P\left( \eta_{id}=\ln 2,\theta=\frac{\pi}{2};\beta\right) =\tfrac{2\sqrt {%
\tanh\left( \hbar\sqrt{\frac{32C_{1}}{17m_{1}}} {\beta}\right)
\tanh\left( \hbar\sqrt{\frac{2C_{1}}{17m_{1}}} {\beta}\right) }}{%
2\tanh\left( \hbar\sqrt{\frac{32C_{1}}{17m_{1}}} {\beta} \right) +%
\frac{1}{2}\tanh\left( \hbar\sqrt{\frac{2C_{1}}{17m_{1}}} {\beta}
\right) }
\end{equation}
which converges to a maximal value at low temperature
\begin{equation}
\lim_{\beta\longrightarrow\infty}P\left( \eta_{id}=\ln2,\theta=\frac{\pi}{2}%
;\beta\right) =\frac{4}{5}.
\end{equation}

\begin{figure}[H]
\centering  \includegraphics[width=12cm,  height=8cm] {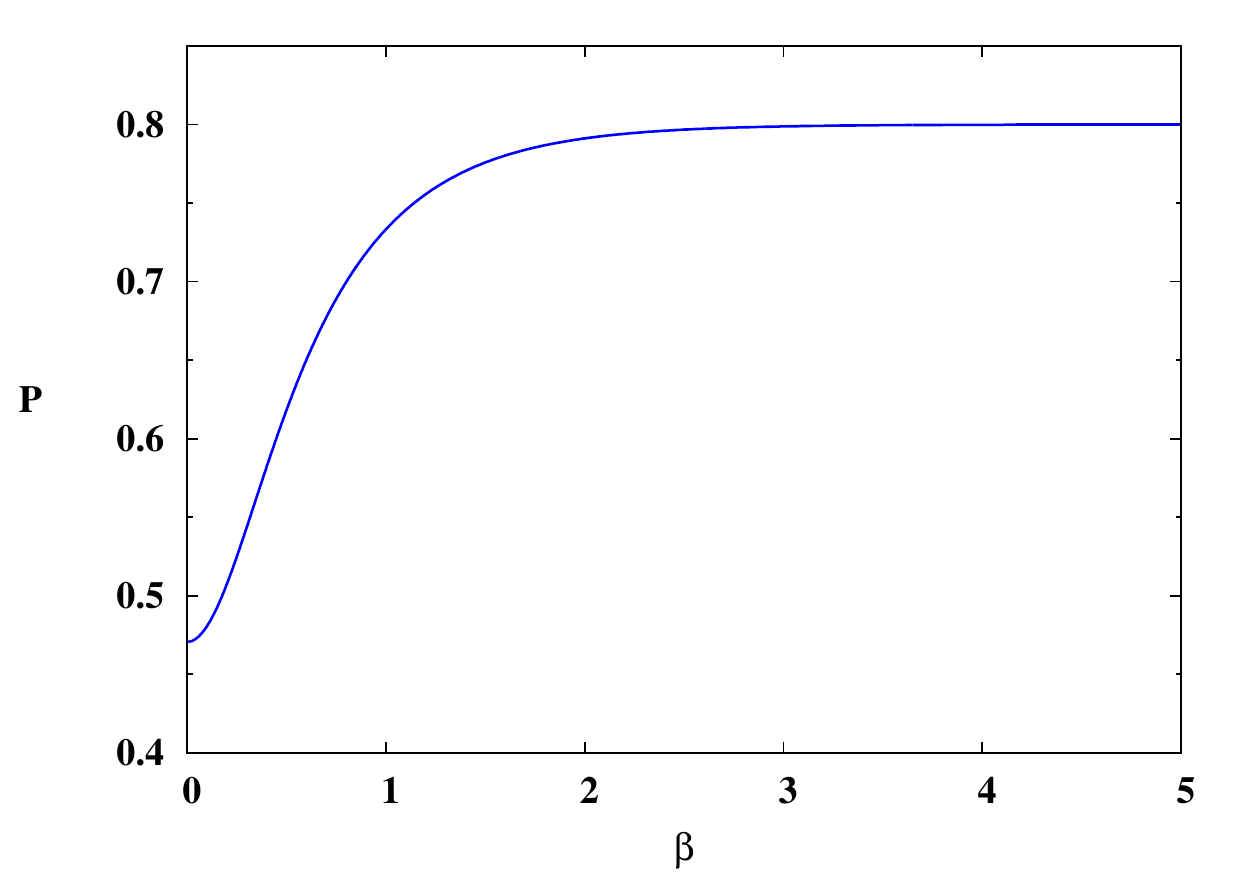} 
\caption{\textsf{Purity versus the temperature for the special values of
$\protect\eta_{id}=
\ln 2$ and 
$\protect\theta=\frac{\protect\pi}{2}$. }}
\label{R46_2}
\end{figure}

We plot the reduced purity in terms of the temperature $\beta$ as  shown
in Figure \ref{R46_2}. One observes that at high temperature the purity is fixed to a
minimal value, as long as the temperature is decreased the purity rapidly
increases to reach a maximal value 0.8 at very low temperature.

\section{Conclusion}


We have analyzed the entanglement of a system
of two coupled harmonic oscillators by using the path integral
mechanism. For this, we have involved a global propagator
based on temperature evolution of our system. Considering
a unitary transformation we were able to explicitly obtain 
the reduced density matrix and therefore the wavefunction describing the whole
spectrum of our system. Later on, we have shown that at high temperature regime
one could obtain the classical distribution that is the Boltzmann probability
and also at low temperature regime the system is in the ground state as should be.

Subsequently, we have used the reduced matrix density to discuss
the entanglement of our system. For this, in the  first stage we have 
calculated the corresponding purity function, which has been obtained 
in terms of different physical parameters. These allowed us
to present different numerical results according to choice of some
configurations of such parameters and therefore extract information
about the system behavior. In fact in different occasions, we have obtained the degree
of the entanglement varying between  maxima and minima values.

We close by mentioning some extension of the present work. Indeed, one could ask about the corresponding
entropies of R\'enyi and von  Neumann. Also, we can ask about the inseparability and the uncertainty relations
by defining the associated Podolsky-Einstein-Rosen operator \cite{LU}. These issues and related matters are actually
under consideration.


\section*{Acknowledgment}


We thank Youness Zahidi for his numerical help.
The authors acknowledge the financial support from King Faisal University.


\end{document}